# Cross-sectional shape analysis for risk assessment and prognosis of patients with true lumen narrowing after type-A aortic dissection surgery

J V Ramana Reddy[1], Toshitaka Watanabe[2,*], Taro Hayashi[3], Hiroshi Suito[1]

[1] Advanced Institute for Materials Research, Tohoku University, 2-1-1 Katahira, Sendai, Miyagi, Japan
[2] Medical Corporation Shimada Clinic Clover Clinic, Osaka, Japan
[3] Department of Cardiovascular Surgery, Akashi Medical Center, Hyogo, Japan

Address corresponding to: Toshitaka Watanabe, MD, PhD
Email: a05m100@gmail.com

## Abstract

**Background:** For acute type-A aortic dissection (ATAAD) surgery, early post-surgery assessment is crucially important for effective treatment plans, underscoring the need for a framework to identify the risk level of aortic dissection cases. We examined true-lumen narrowing during follow-up examinations, collected morphological data 14 days (early stages) after surgery, and assessed patient risk levels over 2.8 years.

**Purpose:** To establish an implementable framework supported by mathematical techniques to predict the risk of aortic dissection patients experiencing true-lumen narrowing after ATAAD surgery.

**Materials and Methods:** This retrospective study analyzed CT data from 21 ATAAD patients. True lumens, false lumens, and full lumens are extracted from the dissected aorta, extending from beyond the left-subclavian-artery bifurcation up to the celiac artery to study postoperative shape changes. Forty uniformly distributed cross-sectional shapes (CSSs) are derived from each lumen to account for gradual changes in shape. We introduced the form factor (FF) to assess CSS morphology. Linear discriminant analysis (LDA) is used for the risk classification of aortic dissection patients. Leave-one-patient-out cross-validation (LOPO-CV) is used for risk prediction.

**Results:** For this investigation, we examined data of 21 ATAAD patients categorized into high-risk, medium-risk, and low-risk cases based on clinical observations of the range of true-lumen narrowing. Our risk classification machine-learning (ML) model used four parameters: the FF of selected CSS obtained from three-lumens, and the centerline (CL) curvature of true lumen at the selected CSS, preserving the model's generalizability. The model's predictions reliably identified low-risk patients, thereby potentially reducing hospital visits. It also demonstrated proficiency in accurately predicting the risk for all high-risk patients.

**Conclusion:** The suggested method anticipates the risk linked to aortic enlargement in patients with a narrowing true lumen in the early stage following ATAAD surgery, thereby aiding follow-up doctors in enhancing patient care.

## Introduction

Aortic dissection involves three terms: true lumen (principal conduit), false lumen (abnormal channel within the aorta wall), and full lumen (combined space of true lumen and false lumen along with the boundary between them). Acute type-A aortic dissection (ATAAD) has a high mortality rate and requires emergency surgery to save lives (1). Emergency surgery is aimed at removing the entry to the false lumen in the ascending aorta. This surgery involves replacing the ascending aorta or ascending to the aortic arch with an artificial vessel. The false lumen often remains after surgery when dissection extends to the descending aorta. Even after entry resection, aortic branches on the false-lumen side (e.g., celiac artery, superior mesenteric artery, renal artery) might act as passageways to the false lumen, leading to enlargement and late stage re-rupture of the remaining descending aorta (2–5). In recent years, the cross-sectional shape (CSS) of the true lumen has attracted attention as a predictor of late enlargement of aneurysms (6, 7). We now know from earlier studies that patients with a narrow true lumen have increased risk of late stage false-lumen enlargement, but not all cases require surgery. Surgery is recommended if the arterial diameter exceeds 55–60 mm or if enlargement is greater than 5 mm in 6 months, posing a high risk of events such as descending aortic rupture. It is unclear how much cases with a narrow true lumen, which are considered to entail high-risk for aortic enlargement, are prone to actual enlargement.

In many patient cases, false-lumen CSSs tend to exhibit a crescent shape, whereas true-lumen CSSs display an elliptical appearance. Several shape analysis studies have used elliptic Fourier analysis (EFA) (8–12), curvature-based approaches (CBA) (13, 14), and various parametric contour approaches (PCA) (15, 16). Bending energy-based approaches derived from the CBA for deformed beams (17) have been used to analyze closed curves in biological and general two-dimensional forms (18–20). However, extending the analysis to a two-dimensional closed curve for classifying similar shapes has some limitations, as observed when classifying similar ellipses of different sizes. Findings indicate EFA as sensitive to noise and shape orientation, whereas CBA and PCA are not exclusively morphological. To address this limitation, we introduced form factor (FF) for morphometric analysis. The method provides a shape-dependent metric that is independent of size and orientation.

This study introduces a mathematical tool to evaluate the deviation from the circular shape of CSS by application of the FF. The findings are expected to contribute to reasonable classification and prediction of patient risk associated with aortic expansion following ATAAD surgery.

## Materials and Methods

### Medical Dataset Construction

#### Dataset Overview

For this study, we adhere to standard clinical acquisition protocols. The CT data collected from 21 retrospectively identified patients with true-lumen narrowing after ATAAD surgery. These patients were followed up for an average of 2.8 years (ranging from 6 months to 9.0 years) at a single institution. The average age at onset was 54.1± 9.1 years old, with 15 male patients (71.4%).

The CT scans are generated from supine-positioned patients in DICOM format. Diagnosis reports are collected securely. Patient-protected health information (PHI) metadata are removed from the DICOM files to ensure patient privacy. All data contributions to this study received approvals from the research ethics committees of Akashi Medical Center (reference number: 30-13) and Tohoku University Hospital (reference number: 2020-1-512).

#### Dataset Collection

Between January 2008 and December 2018, a total of 171 patients underwent ATAAD surgery at Akashi Medical Center. We conducted an analysis of CT scan data. Figure 1(A) presents the categories of these patients during regular check-ups. From these 171 patients, we excluded 58 patients with no CT data available after six months, 2 patients with Marfan syndrome, and 87 patients without true-lumen narrowing of the descending aorta on postoperative CT.

Among the initial 171 patients, 24 exhibited true lumen narrowing, which is defined using a specific ratio: the minor diameter of the true lumen divided by the diameter of the descending aorta. This ratio is required to be less than 0.5. Measurements are taken at the locations of the most severe narrowing. However, three patients were excluded from analyses because of the unavailability of CT data within the initial 14 days (early stages) after ATAAD surgery.

For evaluation, we used a non-ionic contrast agent in contrast-enhanced CT images at various postoperative time points, including early stages, six months, one year, and annually thereafter. However, for the purposes of this study, our focus is explicitly on CT data obtained from the early stages. The early postoperative images covered the region between the aortic arch and aortic bifurcation, with slice thicknesses of 2.5–5.0 mm. These early postoperative CT images are then compared to images taken at six months and annually thereafter.

#### Clinical Classification

The risk classification of 21 patients is established through clinical observation over an average period of 2.8 years. Several guidelines are described in the related literature (21–24). In this article, we selected criteria based on aortic enlargement, with details presented in Table 1.

**Table 1: Clinical risk classification criteria**

| Risk type | Aortic enlargement |
|---|---|
| Low | Less than 5 mm per year (Small expansion) |
| Medium | 5–10 mm per year (Expansion and possible continuation) |
| High | More than 10 mm per year (Rapid expansion and continuation) |

Figure 1(B) presents the proportion of data associated with the risk classification. Figure 1(C) depicts the study's methodology through a detailed flowchart. This methodology is explained further hereinafter.

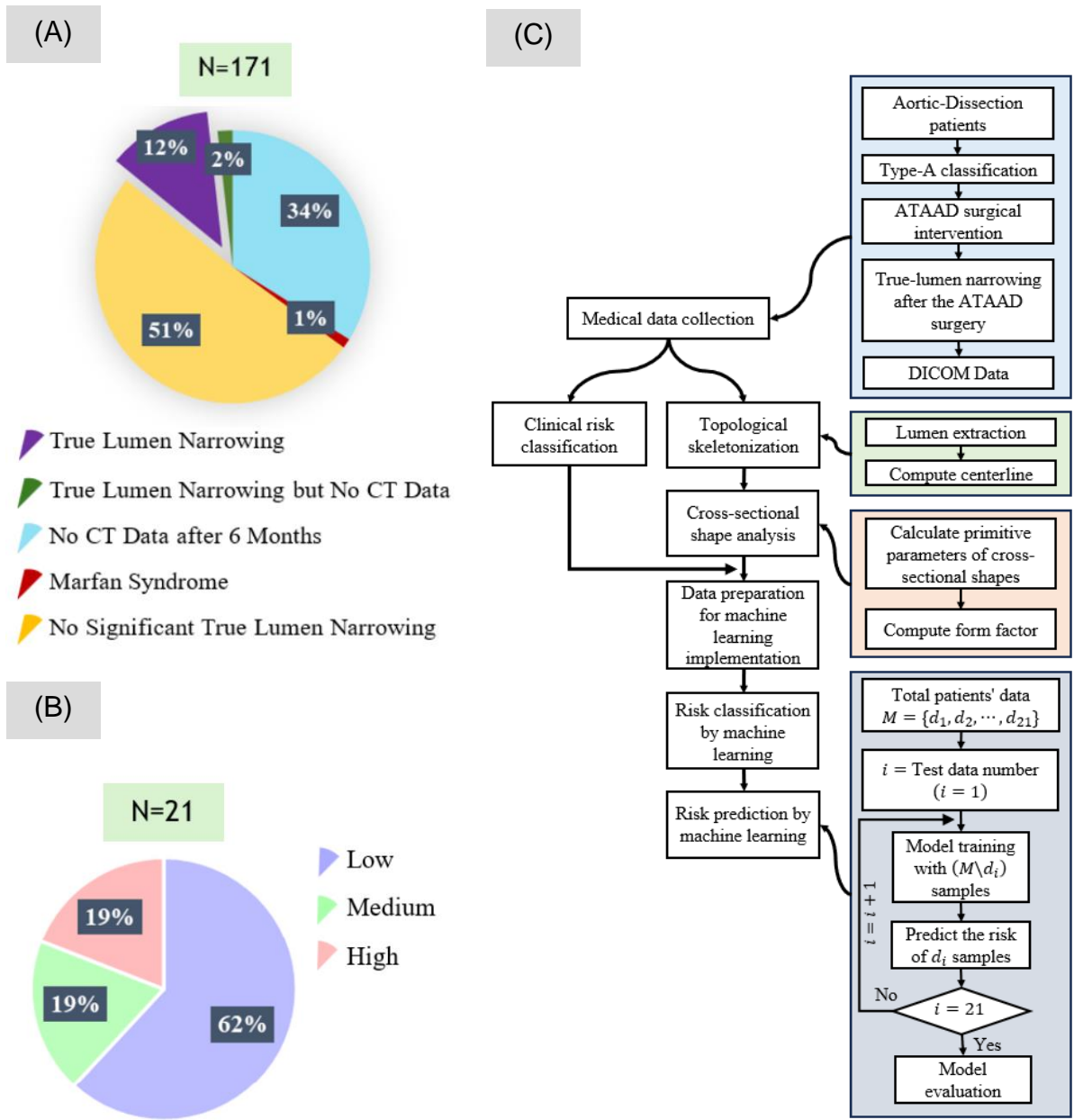

**Figure 1:** (A) Shows categories of clinical data after acute type-A aortic dissection (ATAAD) surgery. (B) Data proportion of risk classification for patients with post-operative narrowing true lumen. (C) Presents the workflow in a flowchart illustration.

## Methodology for deriving Cross-Sectional Shapes

This section outlines the procedure for deriving tangent, normal, binormal (TNB) frame-based CSSs from each aortic-dissected lumen at evenly spaced points along the CLs.

The CT data covered from the aortic arch to the aortic bifurcation. The dissected aorta's true , false , and full lumens are extracted from the left-subclavian-artery bifurcation to the celiac artery. The full lumen is obtained by combining the true lumen and false lumen along with the boundary between them, while extracting true lumen and false lumen is straightforward. These lumen models are derived using ITK-SNAP (25) from CT data of patients diagnosed with ATAAD. They are stored in the Standard Template Library (STL) format, representing the lumen surface with triangulated structures defined by vertices and faces. The CLs of the three lumens are computed using the vascular modeling toolkit (VMTK, www.vmtk.org) library (26). Details of the procedure, the obtained lumens, and their CLs are outlined in Appendix A.

At each evenly spaced point along the CL, the unit tangent vector of the CL is examined. Subsequently, a plane normal to the CL is constructed at that specific point. The normal plane intersection with the lumen's STL model produced a point cloud. The obtained point cloud is refined by ensuring that the distance between any two consecutive points is not too small: specifically, not less than 1 nanometer (1 nm). Consequently, the CSSs are derived from the refined point cloud data within the TNB-frame. A subset of these CSSs is displayed in Figure 2 (A–C).

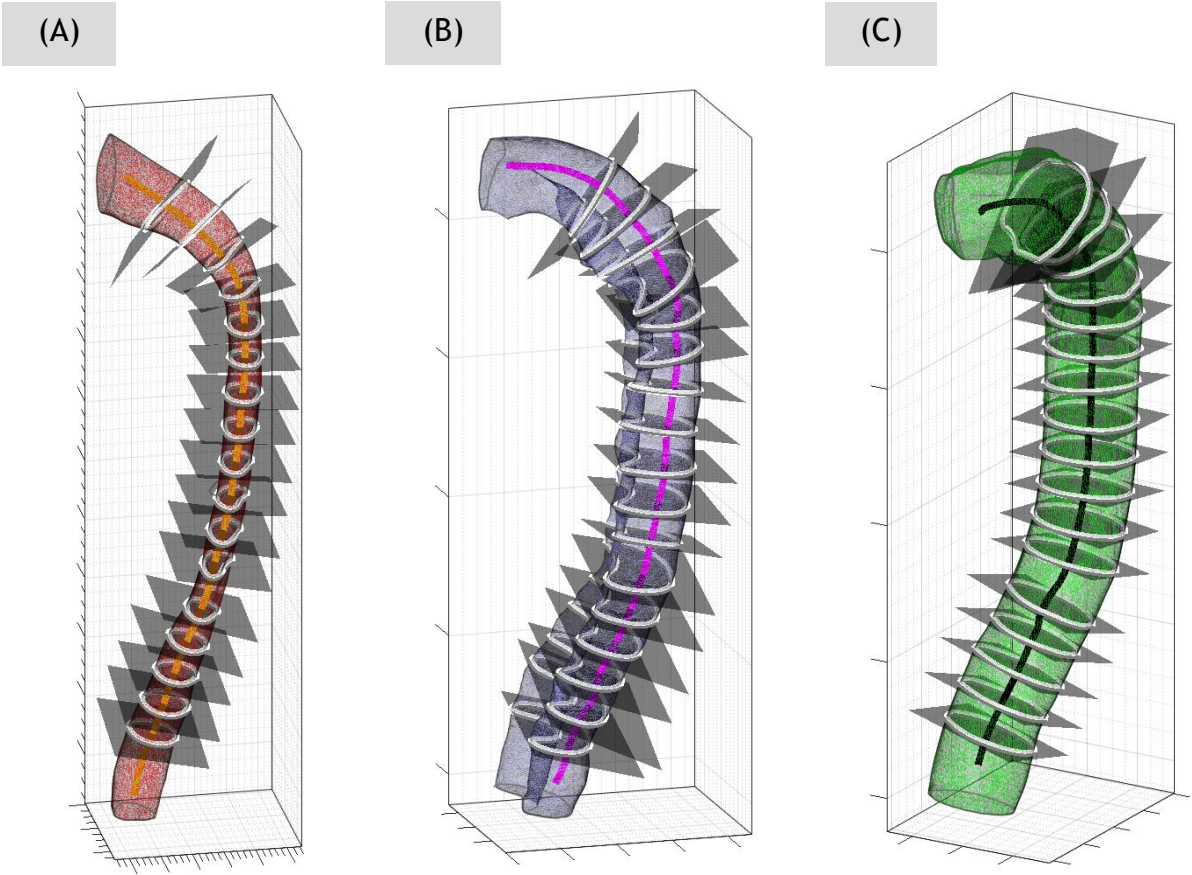

Figure 2: Derived cross-sectional shapes (CSSs) from true, false, and full lumens, along with their corresponding centerlines (CLs), are shown respectively in (A–C).

## Analysis of Cross-Sectional Shapes

As described in the *Introduction*, we sought a formula that can represent the characteristics of crescent-shape CSSs and introduced Form Factor (FF), which is defined as presented below.

> **Definition:** The FF of a closed, simply connected, two-dimensional curve (C) is defined as the ratio of the square of its perimeter length $\bigl(P(C)\bigr)$ to the enclosed area $\bigl(A(C)\bigr)$, as
> $$\mathrm{FF}(C) = \frac{[P(C)]^2}{A(C)}.$$

Considering C as a closed curve described by $n$ points on its boundary, denoted as $\{(x_i, y_i), i = 1, \cdots, n\}$, then the approximate estimation of the area and perimeter are computed as presented below.

Area: The area (A) enclosed by the curve C can be approximated as
$$A(C) = \frac{1}{2}\oint_C x\,dy - y\,dx \approx \frac{1}{2}\sum_{i=1}^{n}(x_i y_{i+1} - x_{i+1} y_i). \qquad (1)$$
This formulation comes from the Green's theorem (27). Green's theorem establishes a relation between the line integral along the shape's boundary and a double integral over the entire area enclosed by the curve.

Perimeter: Perimeter (P) of the closed curve C can be approximated as
$$P(C) = \oint_C \sqrt{\left(\frac{dx}{dt}\right)^2 + \left(\frac{dy}{dt}\right)^2}\,dt \approx \sum_{i=1}^{n}\sqrt{(x_{i+1} - x_i)^2 + (y_{i+1} - y_i)^2}. \qquad (2)$$

Numerical approximations in Eqs. (1) and (2) are based on a discrete set of $n$ points $(x_i, y_i)$ on curve C, where $x_{n+1} = x_1$ and $y_{n+1} = y_1$. It is noteworthy that FF is a scale-independent parameter.

On the one hand, in the civil engineering field, a parameter designated as "resistance to bending" has the same definition as our FF. It represents resistance to bending forces along beam or column axes depending on their CSSs. On the other hand, we use it for a different purpose, representing the risks of expansion in the NB-planes. Examples of FF values for primitive two-dimensional, simply connected closed curves are displayed in Figure 3 as well as for a crescent shape.

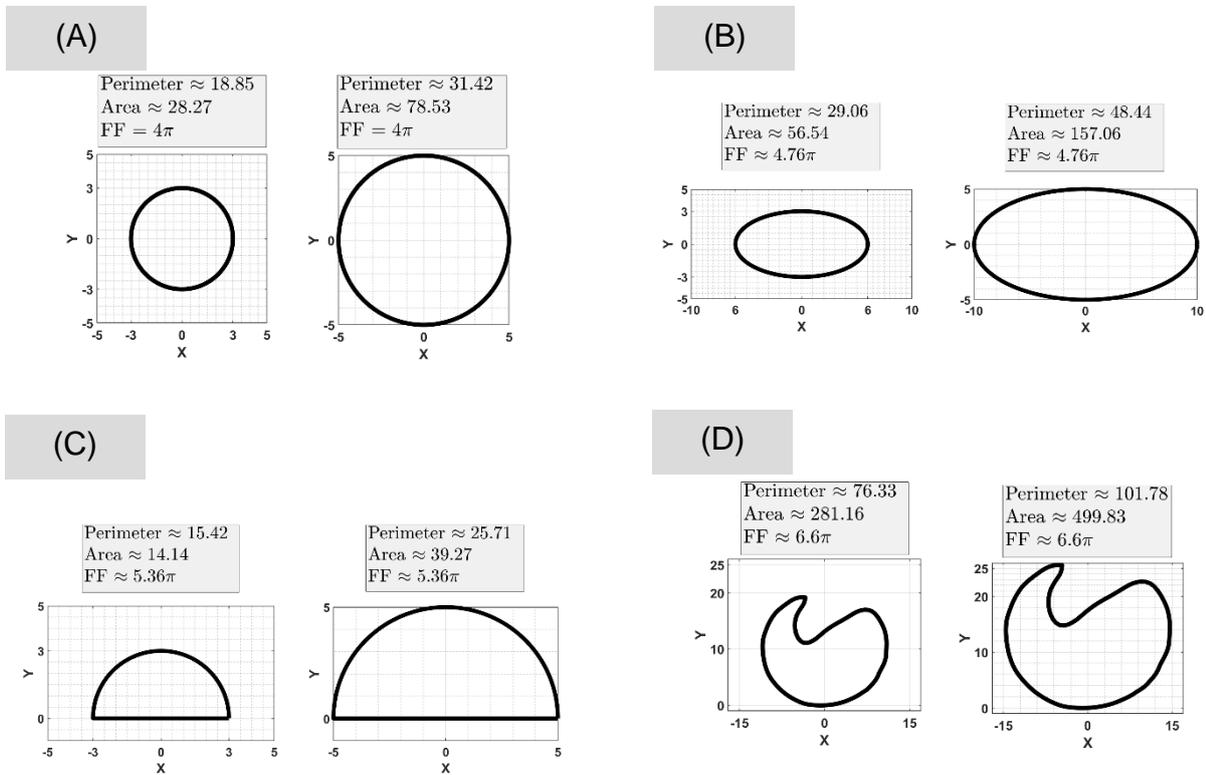

**Figure 3:** The form factors (FFs) of (A) circle, (B) ellipse, (C) semi-circle, and (D) crescent shape.

The FFs are calculated for all 40 CSSs of a false lumen. Among these, the CSS possessing the highest FF has been identified. Subsequently, we computed the FF for corresponding CSS at the same location in true lumen and full lumen. Figure 4 presents the CSS with the highest FF in the false lumen in the NB-plane, along with the corresponding true-lumen and full-lumen CSS for all patients.

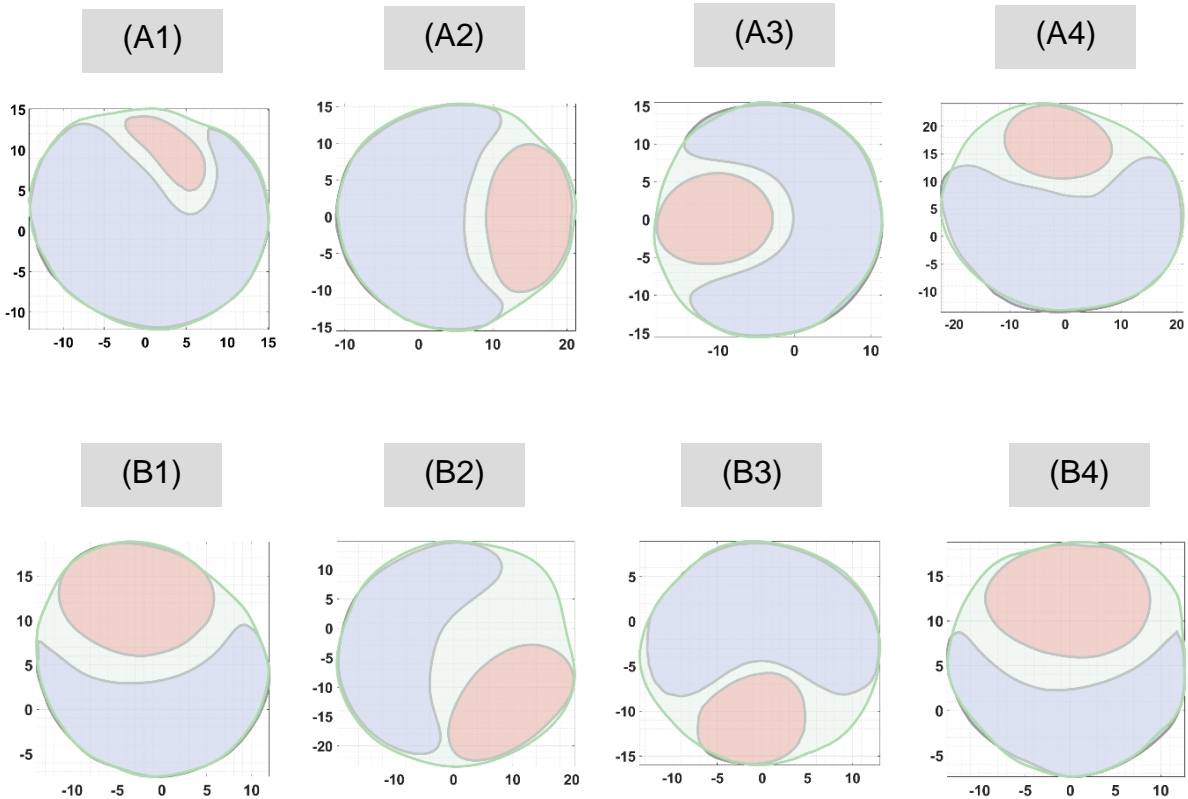

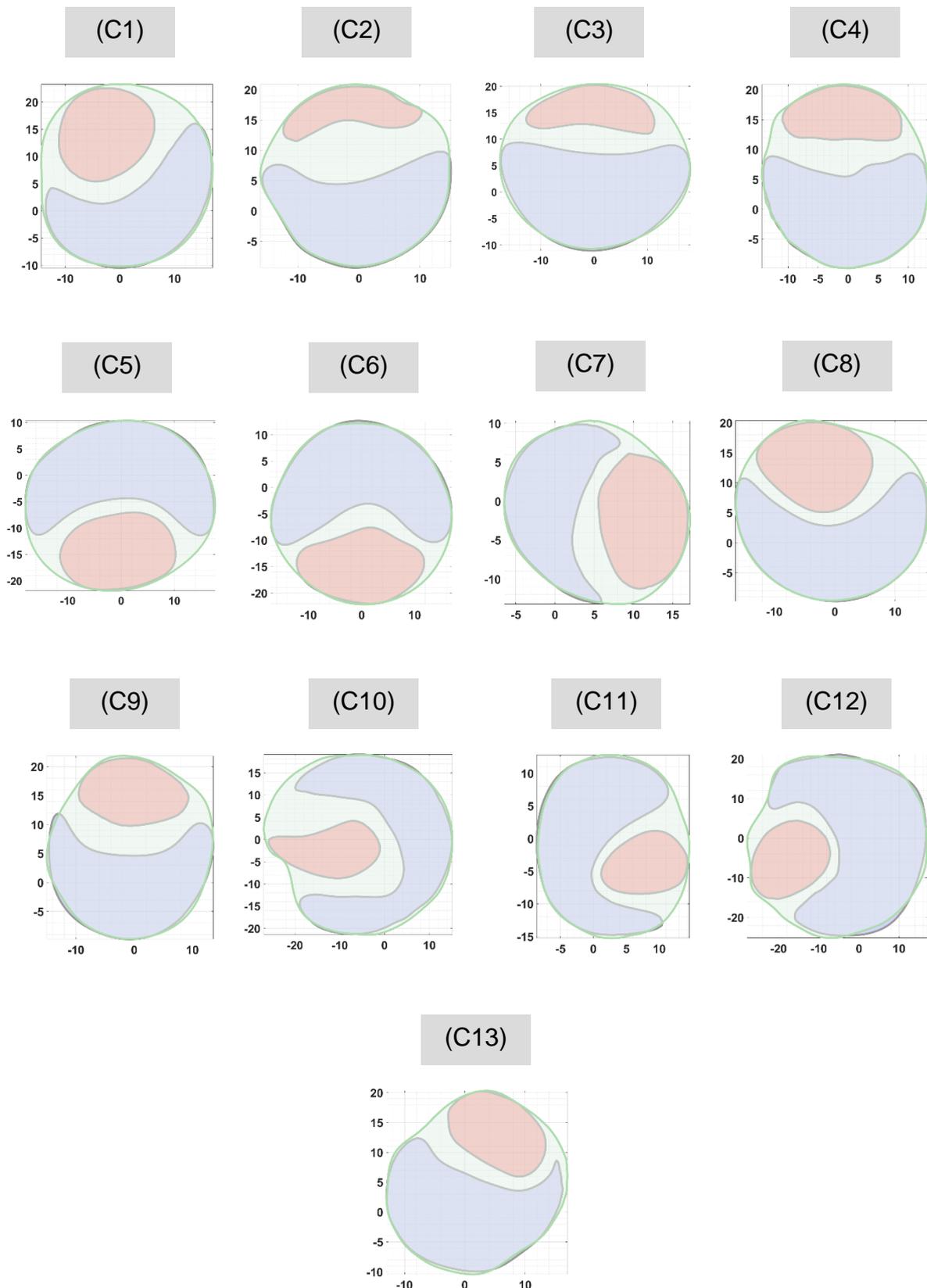

**Figure 4:** CSS of the false lumen in blue, true lumen in red, and full lumen in green for patients categorized as high-risk (A1–A4), medium-risk (B1–B4), and low-risk (C1–C13), at their highest FF (form factor) locations in false lumens.

The dataset has been formed with four derived features: the FF of the CSS, which has the highest FF in the false lumen; the FF of the corresponding true lumen and full lumen at the same location, and the curvature of the true lumen's CL at the same location. These four features are extracted from the CT-scan data and are used to train a ML model.

## Machine Learning Approach

We have adopted LDA for our data analysis because of its effectiveness in distinguishing between classes and providing insights of the dataset. LDA is a supervised dimensionality reduction and classification technique that aims to find a transformation matrix ($W$) for the feature space. Consequently, it reveals a low-dimensional subspace that maximizes separation between classes while minimizing within-class separation (28, 29). The degree of separability is quantified using the Mahalanobis distance (30). The formulation of LDA, which is rooted in statistics, is used in various scientific domains, from biomedical research to pattern recognition. Hereinafter, 'risk' is treated as a 'class' and vice versa.

Because the dataset has a multi-class imbalanced distribution, as presented in Figure 1(B), class weights ($\omega_k$) are adjusted based on class proportions to enhance the model's classification performance and to prevent it from being overly influenced by the majority class.

Consider a dataset $X = [x_1, x_2, \ldots, x_n] \in \mathbb{R}^{(n \times m)}$ with $n$ data points, each with $m$ features, with all entries as real numbers. These data points are labeled into $K$ different imbalanced classes. Let $\mu = \frac{1}{n}\sum_{i=1}^{n} x_i$ be the mean vector of all data points across all classes. Also, let $\mu_k = \frac{1}{n_k}\sum_{i=1}^{n_k} x_i$ be the mean vector of data points in class $k$, which have $n_k$ data points. Then, the weight-adjusted multi-class LDA is definable as shown below.

**Definition:** Let $W = [v_1, v_2, \cdots, v_d]$ be a lower dimensional transformation matrix with columns representing linear combinations of features, i.e., projection subspace ($d << m$). Then, the goal of LDA is to find a linear transformation $W$ that maximizes Fisher's optimization criterion as

$$\mathcal{J}(W) := \frac{W^T \cdot S_B \cdot W}{W^T \cdot S_W \cdot W}.$$

The between-class scatter matrix ($S_B$) and within-class scatter matrix ($S_W$) are computed as

$$\left.\begin{aligned} S_B &= \sum_{k=1}^{K} n_k (\mu_k - \mu) \cdot (\mu_k - \mu)^T \\ S_W &= \sum_{k=1}^{K} \omega_k \left( \sum_{i=1}^{n_k} (x_i - \mu_k) \cdot (x_i - \mu_k)^T \right) \end{aligned}\right\}.$$

Here, $\omega_k \left(:= \frac{n}{n_k \cdot K}\right)$ presents the class weights used to adjust each class's contribution.

Matrix $W$ that maximizes $\mathcal{J}(W)$ will be obtained by differentiating $\mathcal{J}(W)$ with respect to $W$ and by setting the derivative equal to zero.

LDA serves a dual purpose: it functions as a tool for supervised dimensionality reduction and classification, as well as a predictor. For this study, prediction tasks are conducted using the leave-one-patient-out (LOPO) approach, which involves excluding a patient from the model's training dataset and predicting the patient's risk using the trained model. The process is iterated for each patient. Then, the model's performance is evaluated after completing all iterations. The model's risk prediction is cross-validated with the clinical risk data to assess the model's predictive capability. The LOPO-CV approach evaluates a model's generalization ability, preventing it from becoming too patient-specific (31–33). Consequently, the LOPO-CV approach is suited for retrospective studies addressing medical issues. The procedure is illustrated briefly in Figure 5.

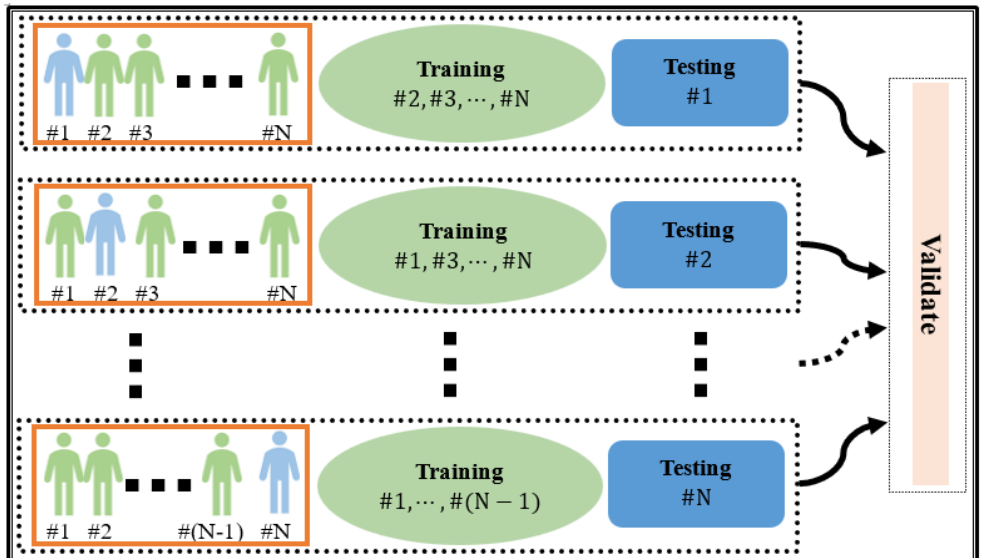

**Figure 5:** Leave-one-patient-out cross-validation (LOPO-CV) approach illustration using patients' data.

# Results

Following the description in the preceding section, we use the LOPO-CV approach to assess the predictive capability of weight-adjusted multi-class LDA. The decision boundaries are computed for all LOPO-CV cases. Figure 6 presents the patient's risk prediction along with the decision boundaries for high-risk cases, whereas the remaining medium-risk and low-risk cases are presented in Appendix B. It is readily apparent that all high-risk cases are classified correctly.

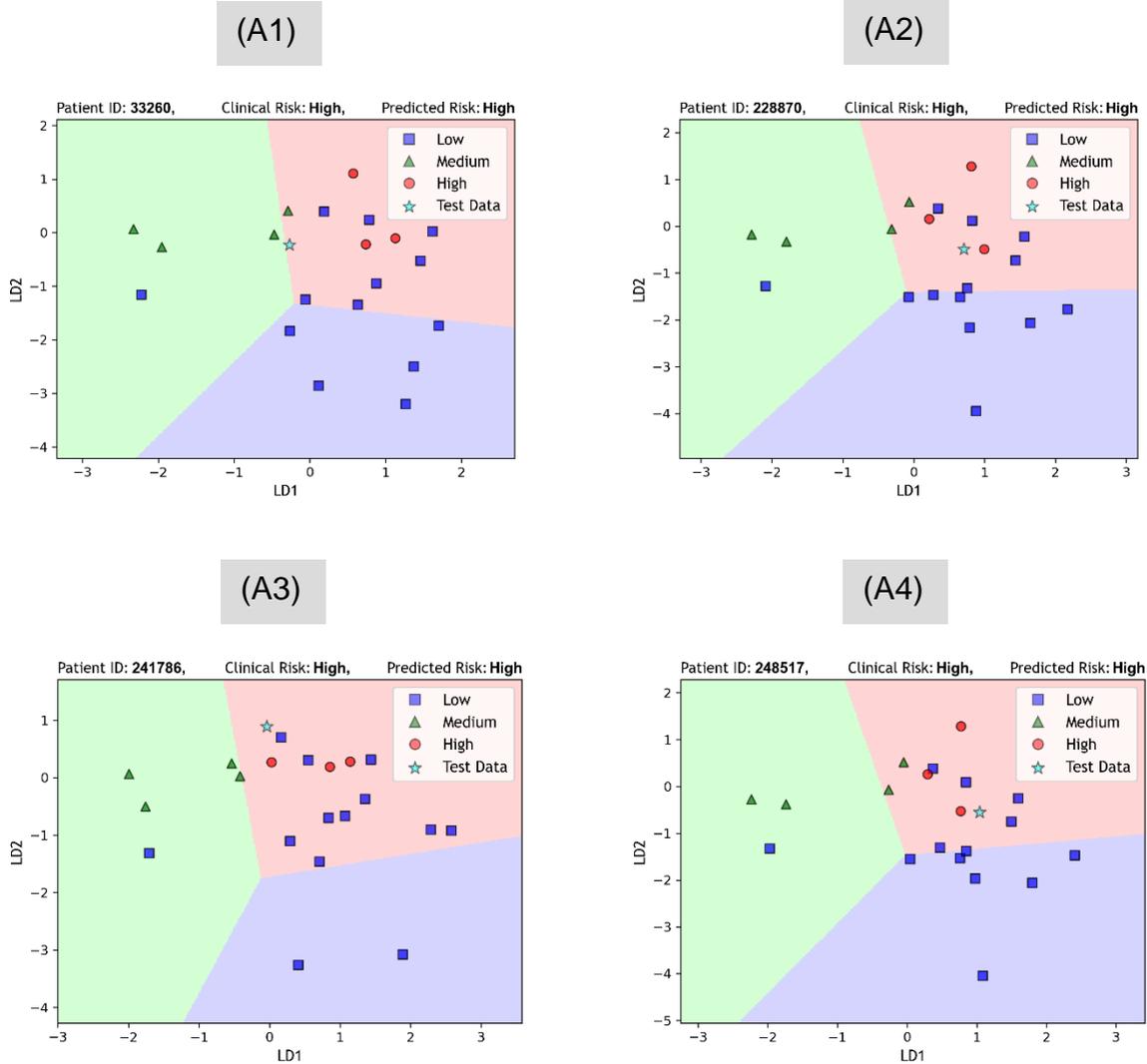

**Figure 6:** Test data risk prediction and associated decision boundaries generated by weight-adjusted multi-class linear discriminant analysis (LDA) for high-risk (A1–A4) cases in leave-one-patient-out cross-validation (LOPO-CV).

The predictive performance of LDA is then assessed using the confusion matrix (Table 2(A)) and relevant metrics. In addition, class-specific confusion matrices for low-risk, medium-risk, and high-risk categories are shown in Table 2(B–D).

For evaluating model performance on multi-class imbalanced data through class-specific confusion matrices, metrics using both rows, such as accuracy $\left(\text{defined as } \frac{TP+TN}{TP+TN+FP+FN}\right)$ and geometric mean $\left(\text{defined as } \sqrt{\frac{TP}{TP+FN} \times \frac{TN}{TN+FP}}\right)$, are found to be inadequate for addressing changes in class distribution (34, 35). Several metrics have been designed to evaluate model predictions in multi-class imbalanced data, including True Positive Rate (TPR or Sensitivity), True Negative Rate (TNR or Specificity), False-Positive Rate (FPR), False-Negative Rate (FNR), Precision (Positive Predictive Value, PPV), and Negative Predictive Value (NPV). Formulations for deriving these metrics and their numerical values for the provided three-class confusion matrices (Table 2(B–D)) are presented in Table 3.

**Table 2: Multi-class confusion matrix (A) and class-specific confusion matrices for risk categories, delineating low-risk (B), medium-risk (C), and high-risk (D) assessments**

(A)

| Clinical Risk | Total 21 | Predicted Risk | | |
|---|---|---|---|---|
| | | Low 6 | Medium 4 | High 11 |
| Low | 13 | 6 | 1 | 6 |
| Medium | 4 | 0 | 3 | 1 |
| High | 4 | 0 | 0 | 4 |

(B)

| Clinical Risk | Predicted Risk | |
|---|---|---|
| | Low | Medium or High |
| Low | 6 $TP_L$ | 7 $FN_L$ |
| Medium or High | 0 $FP_L$ | 8 $TN_L$ |

(C)

| Clinical Risk | Predicted Risk | |
|---|---|---|
| | Medium | Low or High |
| Medium | 3 $TP_M$ | 1 $FN_M$ |
| Low or High | 1 $FP_M$ | 16 $TN_M$ |

(D)

| Clinical Risk | Predicted Risk | |
|---|---|---|
| | High | Low or Medium |
| High | 4 $TP_H$ | 0 $FN_H$ |
| Low or Medium | 7 $FP_H$ | 10 $TN_H$ |

**Table 3: Evaluation metrics for analyzing the class-specific confusion matrices in cases involving imbalanced low-risk, medium-risk, and high-risk classes**

| Evaluation Metrics | Low | Medium | High |
|---|---|---|---|
| $TPR = \left(\frac{TP}{TP+FN}\right)$ | 0.46 | 0.75 | 1.00 |
| $TNR = \left(\frac{TN}{TN+FP}\right)$ | 1.00 | 0.94 | 0.59 |
| $FPR = 1 - TNR$ | 0.00 | 0.06 | 0.41 |
| $FNR = 1 - TPR$ | 0.54 | 0.25 | 0.00 |
| $PPV = \left(\frac{TP}{FP+TP}\right)$ | 1 | 0.75 | 0.36 |
| $NPV = \left(\frac{TN}{FN+TN}\right)$ | 0.53 | 0.94 | 1 |

## Discussion

In accordance with the three-class confusion matrix (Table 2(A)), the true positives (TP) encompassing 6 low-risk patients, 3 medium-risk patients, and 4 high-risk patients are correctly identified by the model. All values in the lower diagonal entries of the multi-class confusion matrix (Table 2(A)) are zeros, which indicates that the model avoided misclassifying higher-risk patients as lower-risk, which is a crucially important aspect of practical applications.

The model exhibits high accuracy in identifying high-risk patients, which can facilitate timely medical intervention. The model reliably distinguishes low-risk patients. In fact, when the model predicts that a patient is at low risk, it accurately confirms the risk status, thereby minimizing unnecessary hospital visits and optimizing resource allocation. Identifying patients with a medium-risk profile is a complex task. The model incorrectly predicted some patients as high-risk when they were actually at low risk. This prediction can engender unnecessary medical interventions for these patients.

The class-specific (two-class) confusion matrices in Table 2(B–D) and metrics in Table 3 provide another evaluation of the model's performance across distinct classes. For the low-risk cases in Table 2(B), the model achieved high specificity (1.00), no false positives (0.00), and perfect precision (1.00), but a moderate false-negative rate (0.54) shows some missed low-risk instances. In medium-risk cases in Table 2(C), the model demonstrated good sensitivity (0.75) and specificity (0.94), with small false positives (0.06) and a moderate false-negative rate (0.25), indicating one missed medium-risk instance.

For high-risk cases in Table 2(D), the model achieved perfect sensitivity (1.00) but showed some difficulty with specificity (0.59). There are no false negatives, demonstrating the model's strength in not missing actual high-risk cases.

Overall, the LDA model accurately predicts high-risk patients and consistently confirms the low-risk status when predicting a patient's risk as low. Some mispredictions, particularly a few low-risk patients being incorrectly labeled as high-risk, have been observed. Improving predictive accuracy appears feasible through augmentation of the training dataset with a larger cohort of patients.

It must be emphasized that the model relied on only four derived features for risk prediction, such a small number of parameters chosen to enhance generalization ability. In fact, we tested the use of more features, which demonstrated that the classifier performance for each case was improved. However, this improvement came at the cost of a loss in generalization ability, which led to poor results in LOPO-CV. Such a phenomenon is known as overfitting. Therefore, this study revealed that the model's generalization ability is much improved when using a rather small number of features.

## Conclusion

The findings of this study suggest is that employing FF of selected CSSs data in LDA effectively predicts the degree of future aortic enlargement in patients with narrowing true lumen after acute type-A dissection surgery.

# Appendix A

For 3D medical image segmentation, we used ITK-SNAP software, which offers active contour segmentation, manual delineation, and image navigation capabilities[1].

In the extracted lumen models, the VMTK[2,3] library is employed to compute the CLs for the full lumen, false lumen, and true lumen. At the outset, the non-uniformly spaced points of the CLs are evenly spaced by enabling the standard procedure. The Frenet–Serret framework, a mathematical tool, is used to analyze a curve's curvature and orientation properties in three-dimensional space.

Letting $s$ denote the arc length parameter and letting $\vec{r}$ represent the position vector of $s$, then tangent vector $(\vec{T})$ signifies the direction of the curve at any given point, indicating the instantaneous direction. It is calculated as the derivative of the position vector with respect to $s$, denoted as $\vec{T} = d\vec{r}/ds$. Then it is normalized to obtain the unit tangent vector $\frac{d\vec{r}/ds}{\|d\vec{r}/ds\|}$.

The normal vector $(\vec{N})$, which points in the direction of curvature at each point on the curve, is given by the derivative of $\vec{T}$ with respect to $s$, denoted as $\vec{N} = \frac{d\vec{T}}{ds}$. The binormal vector, $\vec{B}$, which is perpendicular to both $\vec{T}$ and $\vec{N}$, describes the orientation of the curve in three-dimensional space. It is calculated as $\vec{B} = \vec{T} \times \vec{N}$.

The Frenet–Serret equations governing the evolution of the tangent, normal, and binormal vectors along the curve are the following:

$$\frac{d\vec{T}}{ds} = \kappa \cdot \vec{T},$$

$$\frac{d\vec{N}}{ds} = -\kappa \cdot \vec{T} + \tau \cdot \vec{B},$$

$$\frac{d\vec{B}}{ds} = -\tau \cdot \vec{N},$$

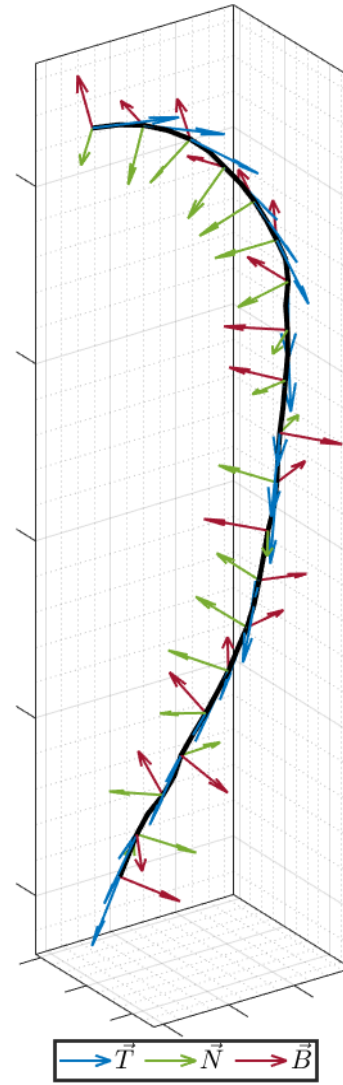

**Figure A1:** Orthonormal tangent $(\vec{T})$, normal $(\vec{N})$, and binormal $(\vec{B})$ vectors are computed at evenly spaced points along the CL (centerline).

where $\kappa$ stands for the curvature and $\tau$ denotes the torsion.

This framework yields three essential unit orthonormal vectors and determines the curvature and torsion at each point on the CL. The orthonormal vectors at evenly spaced locations on the CL are presented in Figure A1. This information represents the lumen's central skeletal structure, which retains its geometric properties, but it omits intricate surface details. Figure A2 displays the true lumens and false lumens along with their CLs for all 21 patients.

---

[1] Yushkevich PA, Piven J, Hazlett HC, Smith RG, Ho S, Gee JC, Gerig G. User-guided 3D active contour segmentation of anatomical structures: significantly improved efficiency and reliability. Neuroimage 2006;31(3):1116–1128.

[2] Antiga L, Piccinelli M, Botti L, Ene-Iordache B, Remuzzi A, Steinman DA. An image-based modeling framework for patient-specific computational hemodynamics. Medical & biological engineering & computing 2008;46: 1097–1112.

[3] Izzo R, Steinman D, Manini S, Antiga L. The vascular modeling toolkit: a python library for the analysis of tubular structures in medical images. Journal of Open Source Software 2018;3(25):745.

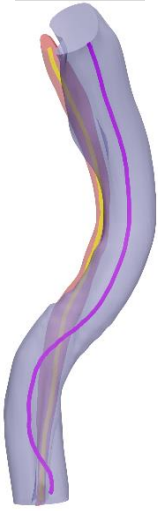 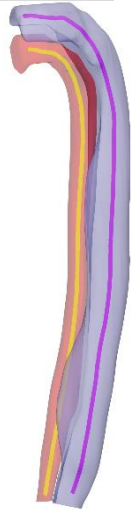 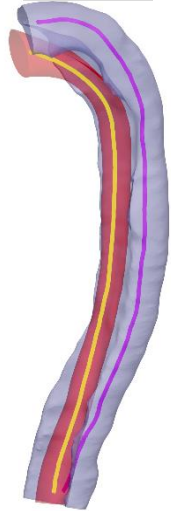 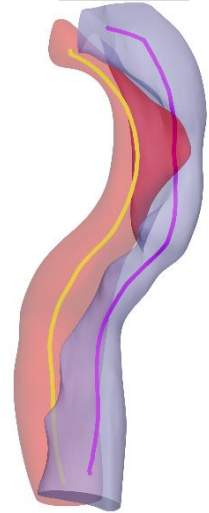
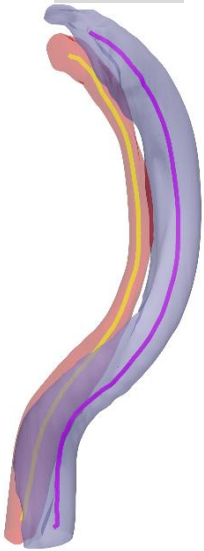 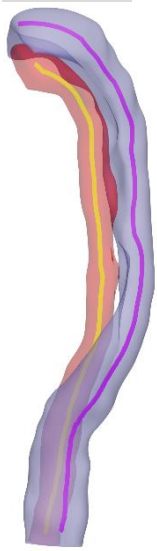 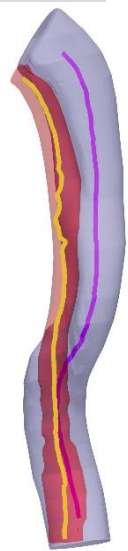 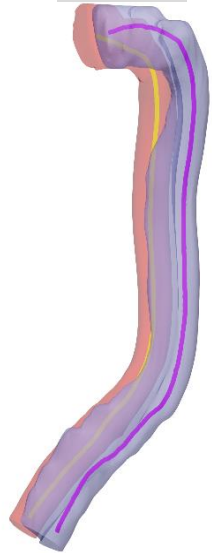
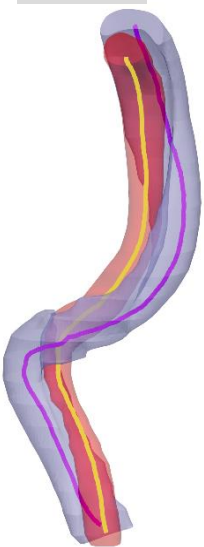 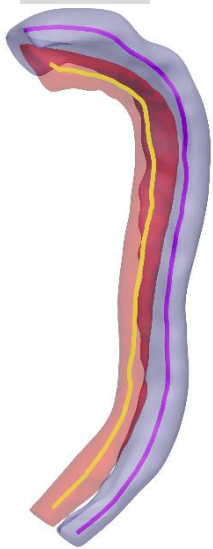 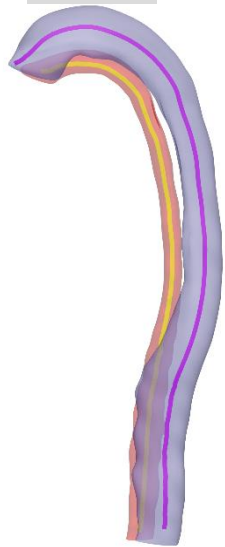 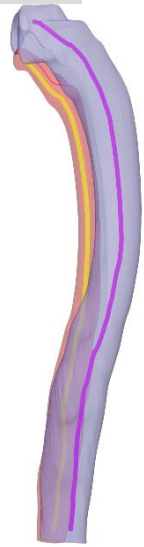

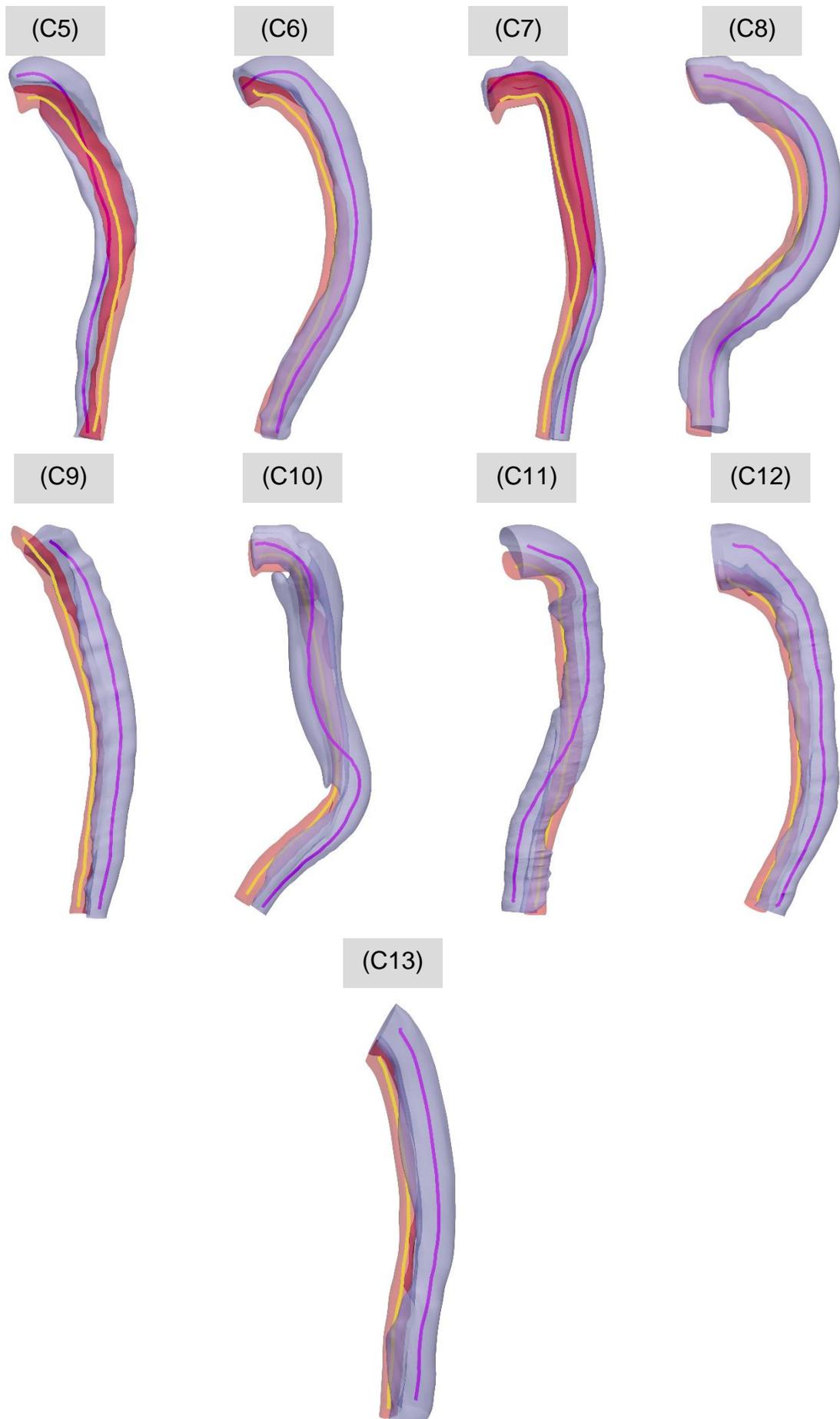

**Figure A2:** True lumen in red and false lumen in lilac, along with their centerlines, for patients clinically categorized as high-risk (A1–A4), medium-risk (B1–B4), and low-risk (C1–C13).

# Appendix B

Figure B1 presents scatterplots obtained using the weight-adjusted multi-class LDA with LOPO-CV, depicting decision boundaries and predicted risk for patient data in the medium-risk **(B1–B4)**, and low-risk **(C1–C13)** categories during the LOPO-CV process.

## Medium risk

(B1)

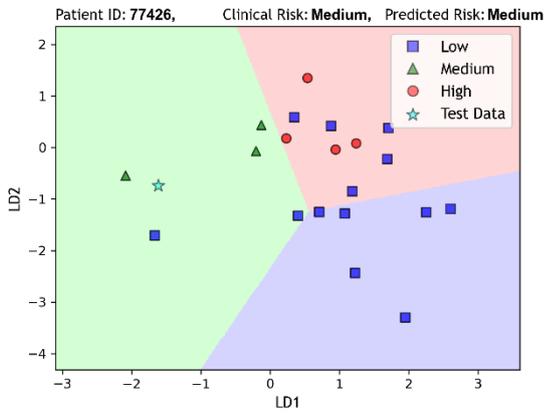

(B2)

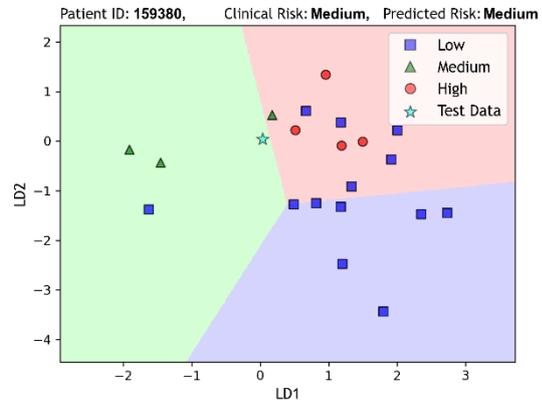

(B3)

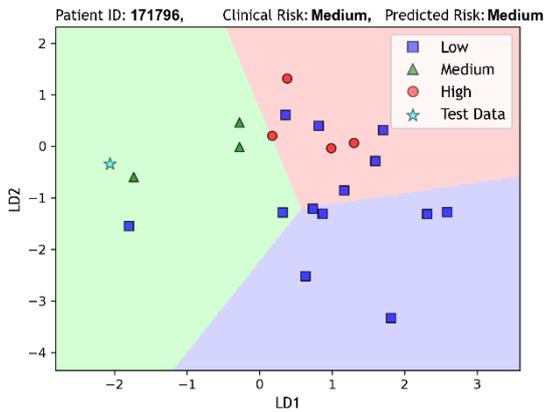

(B4)

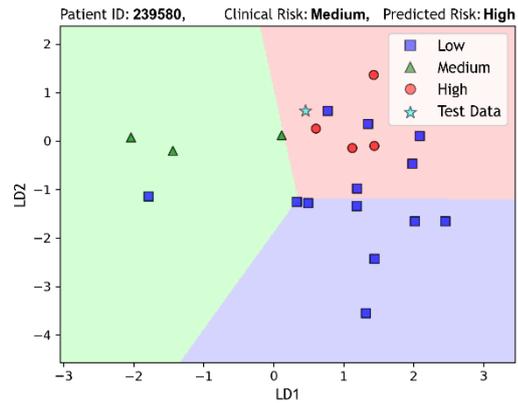

## Low risk

(C1)

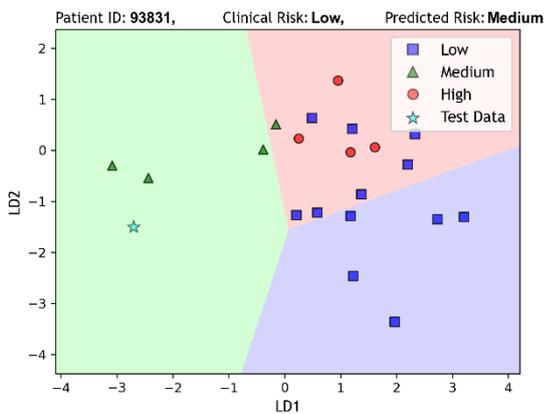

(C2)

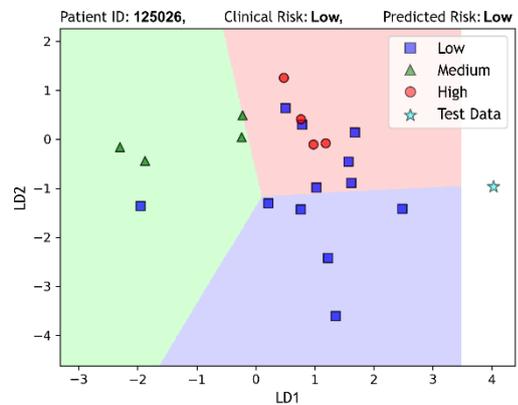

(C3)

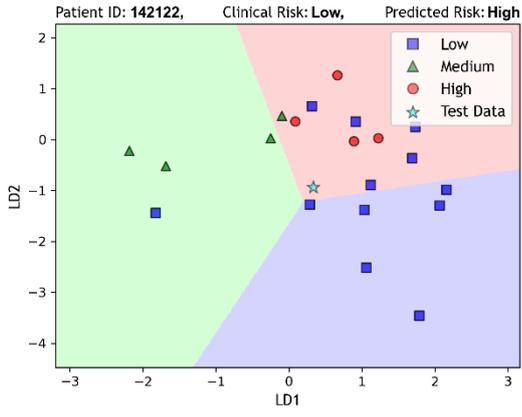

(C4)

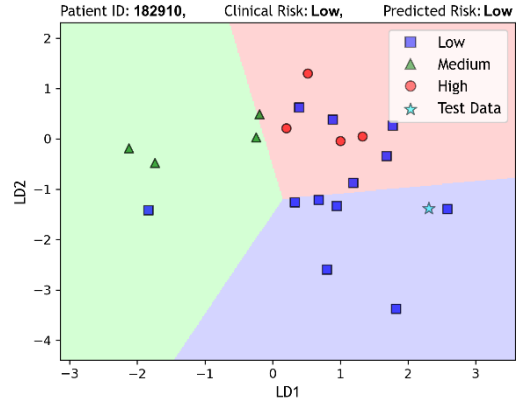

(C5)

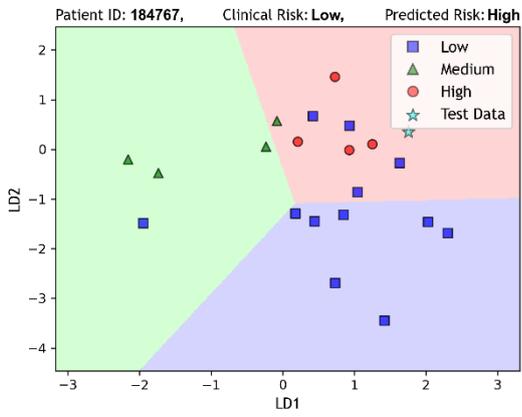

(C6)

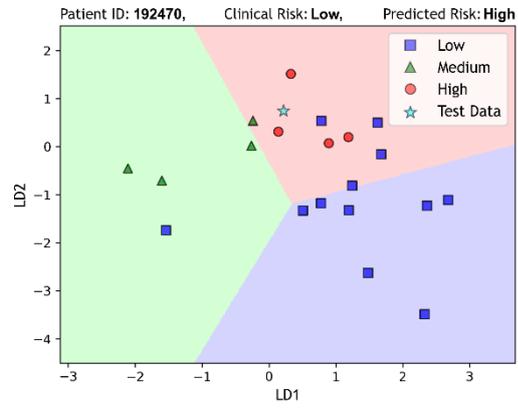

(C7)

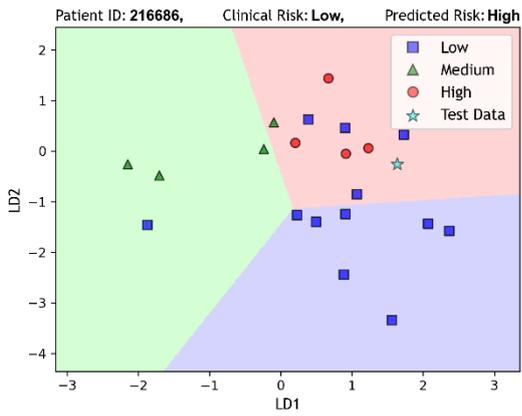

(C8)

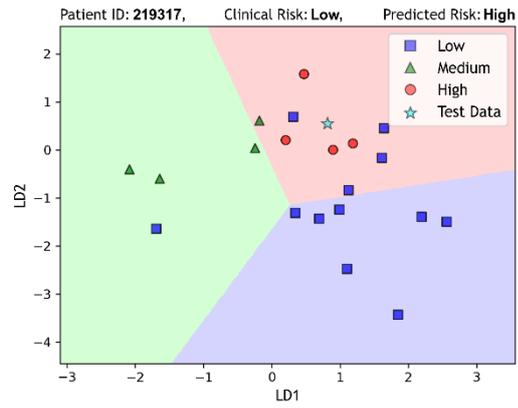

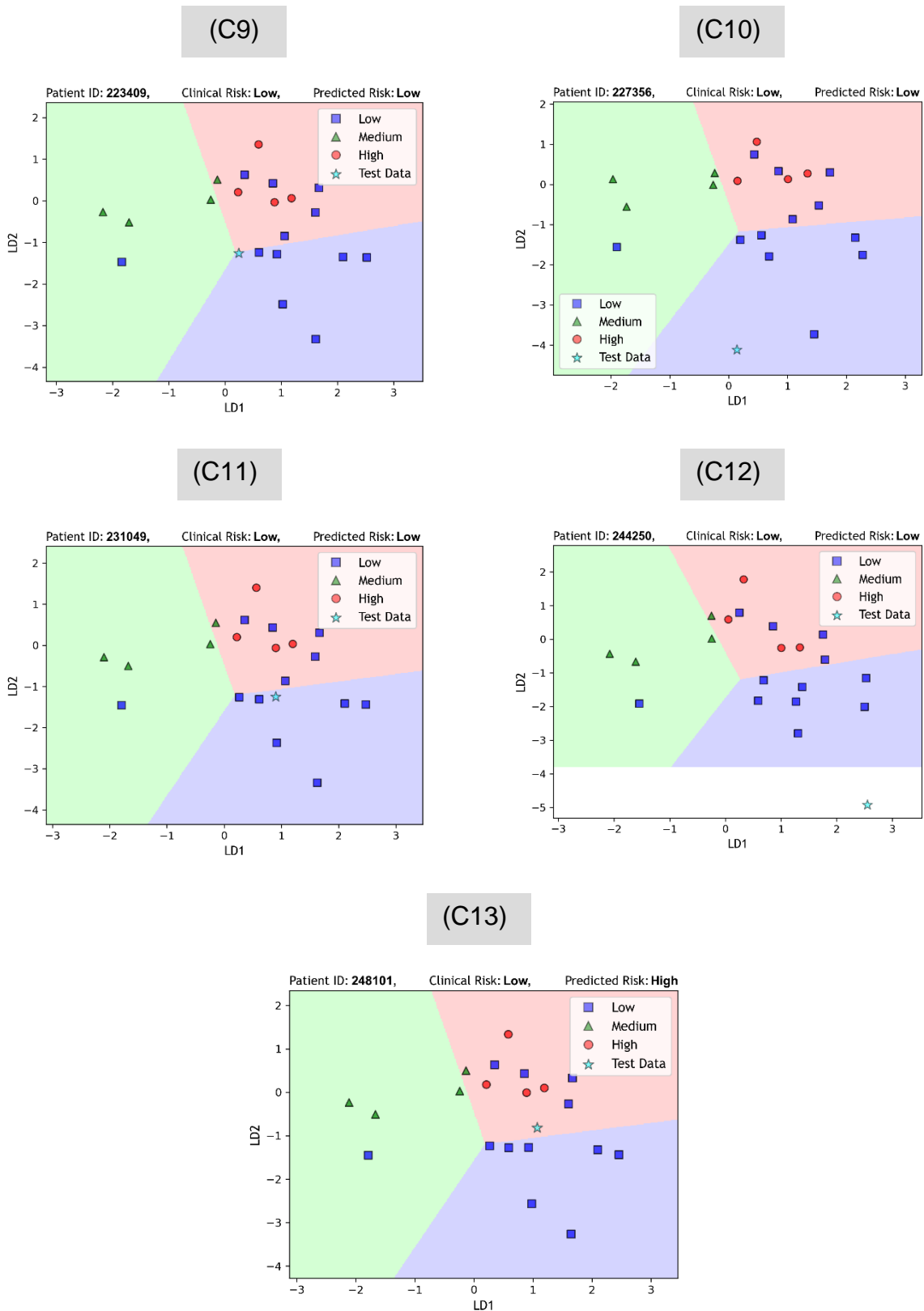

**Figure B1:** Test case risk prediction and associated decision boundaries generated by weight-adjusted multi-class linear discriminant analysis (LDA) for medium-risk (B1–B4) and low-risk (C1–C13) cases in leave-one-patient-out cross-validation (LOPO–CV).